# Title: Liquid crystal-enabled electroosmosis through spatial charge separation in distorted regions as a novel mechanism of electrokinetics

**Authors:** Israel Lazo, Chenhui Peng, Jie Xiang, Sergij V. Shiyanovskii, and Oleg D. Lavrentovich*

**Affiliation:**

Liquid Crystal Institute and Chemical Physics Interdisciplinary Program, Kent State University, Kent, OH 44242, USA

*olavrent@kent.edu

**Abstract**: Electrically-controlled dynamics of fluids and particles at microscales is a fascinating area of research with applications ranging from microfluidics and sensing to sorting of biomolecules.  The driving mechanisms are electric forces acting on spatially separated charges in an isotropic medium such as water.  Here we demonstrate that anisotropic conductivity of liquid crystals enables new mechanism of highly efficient electro-osmosis rooted in space charging of regions with distorted orientation.  The electric field acts on these distortion-separated charges to induce liquid crystal-enabled electro-osmosis (LCEO).  LCEO velocities grow with the square of the field, which allows one to use an AC field to drive steady flows and to avoid electrode damage.  Ionic currents in liquid crystals that have been traditionally considered as an undesirable feature in displays, offer a broad platform for versatile applications such as liquid crystal enabled electrokinetics, micropumping and mixing.

Electrically driven flows of fluids (electro-osmosis) and transport of particles (electrophoresis), collectively called electrokinetics, represent an active area of research in science of soft matter and in development of microfluidic, electrochemical and biomedical devices[1,2]. Research and applications focus on isotropic electrolytes such as water. Electro-osmosis in water is of two types, broadly classified as linear and nonlinear. In the linear effect, the driving electric field $E$ acts on the charges forming an electric double layer at the solid/fluid interface. The ions in the double layer drag the fluid with the typical velocity growing linearly with the applied field, $u_{linear} = -\varepsilon\varepsilon_0 \varsigma E/\eta$, where $\varepsilon$ and $\eta$ are the dielectric constant and viscosity of water, respectively, $\varepsilon_0$ is the electric constant, and $\varsigma$ is the constant field-independent zeta-potential characterizing the fixed surface charges. A nonlinear electro-osmosis is best described by the so-called induced charge electro-osmosis (ICEO) around a conductive (metallic) particle of a radius $a$ placed in water[3,4,5]. The applied electric field induces ionic currents in water. Since the ions cannot penetrate the particle, they create double layers with the polarization charges of the particle. The field drags the charges around the particle's surface to set a quadrupolar ICEO flow. The ICEO velocity is proportional to $E^2$, since the induced zeta potential scales as $aE$: $u_{cond} = -\varepsilon\varepsilon_0 aE^2/\eta$, Ref.[5]. If the particle is not conductive (say, a dielectric sphere), the induced zeta potential is weak, $\varsigma \sim \lambda_D E$, where $\lambda_D \ll a$ is the Debye screening length, typically on the scale of nanometers; the resulting velocities are small, $u_{diel} = -\varepsilon\varepsilon_0 \lambda_D E^2/\eta$, Ref.[3,4].

In this work, we demonstrate a new mechanism of electro-osmosis based on formation of a space charge through intrinsic anisotropy of conductivity of a liquid crystal (LC) used as an electrolyte. When the LC orientation, specified by the so-called director $\hat{\mathbf{n}}$, varies in space, application of an electric field leads to separation of charges. The space charge build-up in homogeneous LCs, known as the Carr-Helfrich effect, has been studied in a context of field-induced anomalous re-alignment and hydrodynamic instabilities [6]. It is considered as an undesirable response of an LC to an electric field in mainstream application, LC displays. In the context of electrokinetics, however, the induced space charge gives rise to the LC enabled electro-osmosis (LCEO) with features that are highly desirable in microfluidic applications. Among these are efficient charge separation at practically any spatial scale without involvement of polarizable surfaces or particles, control of electro-osmotic flows and electrophoretic transport by LC alignment or through dynamic changes of this alignment, possibility to use AC field as a driving force for sustainable flows and transport. The direction of electroosmotic flows in LC is shown to depend on the type of director distortions. For example, a spherical inclusion that forces the liquid crystal director to be tangential to its surface causes an electroosmotic flow in direction that is opposite to the flows around a similar spherical inclusion with perpendicular director alignment. Last but not least, the microfluidic motion in LCEO does not require mechanical drivers such as pumps or capillary action.

## Results

**Charge separation and electro-osmotic flows in a nematic.** In a classic ICEO effect observed in isotropic electrolytes, charge separation is caused by the conductive nature of the metallic particle, Fig.1a. In a liquid crystal, the mechanism of spatial separation of charges does not rely on the particle's properties and instead is rooted in anisotropy of conductivity and director distortions of the anisotropic electrolyte, Fig.1b-d. The ground state of the simplest LC, the so-called uniaxial nematic, is a uniform alignment of molecules along a single direction, described by the director $\hat{\mathbf{n}}_0 = const$. Suppose that the director is distorted. These distortions might be caused, for example, by a spherical particle placed in the otherwise uniform nematic.

Anisotropy of molecular interactions sets a preferred orientation of the director at the particle's surface that can be perpendicular, Fig.1b, or tangential, Fig.1c, to this surface. The particle-induced distortions $\hat{\mathbf{n}}(\mathbf{r})$ relax to the uniform state $\hat{\mathbf{n}}_0 = const$ over a length scale comparable to the sphere's radius $a$ [7]. To stress the difference with the ICEO, we consider a particle that is not conductive. LCs are anisotropic electrolytes, with the electric conductivity $\sigma_\parallel$ along $\hat{\mathbf{n}}$ different from the conductivity $\sigma_\perp$ perpendicular to it. Typically, $\Delta\sigma = \sigma_\parallel - \sigma_\perp > 0$. In the electric field $\mathbf{E}$, the ions separate and create clouds of space charge, according to the director pattern $\hat{\mathbf{n}}(\mathbf{r})$. For example, in Fig.1b, the director lines converge on the left and right-hand sides of the particle. These regions become enriched with the positive and negative charges, respectively, for the shown polarity of $\mathbf{E}$ and $\Delta\sigma > 0$. The induced charge clouds experience a bulk force of the applied field, which sets up the LCEO flows. The pattern of these flows, in particular, their polarity, is determined by the director configuration $\hat{\mathbf{n}}(\mathbf{r})$. For example, in Fig.1b, one expects flows of a "puller" type with the horizontal $x$-streams directed *towards* the sphere, while in Fig.1c, the polarity is of the opposite "pusher" type.

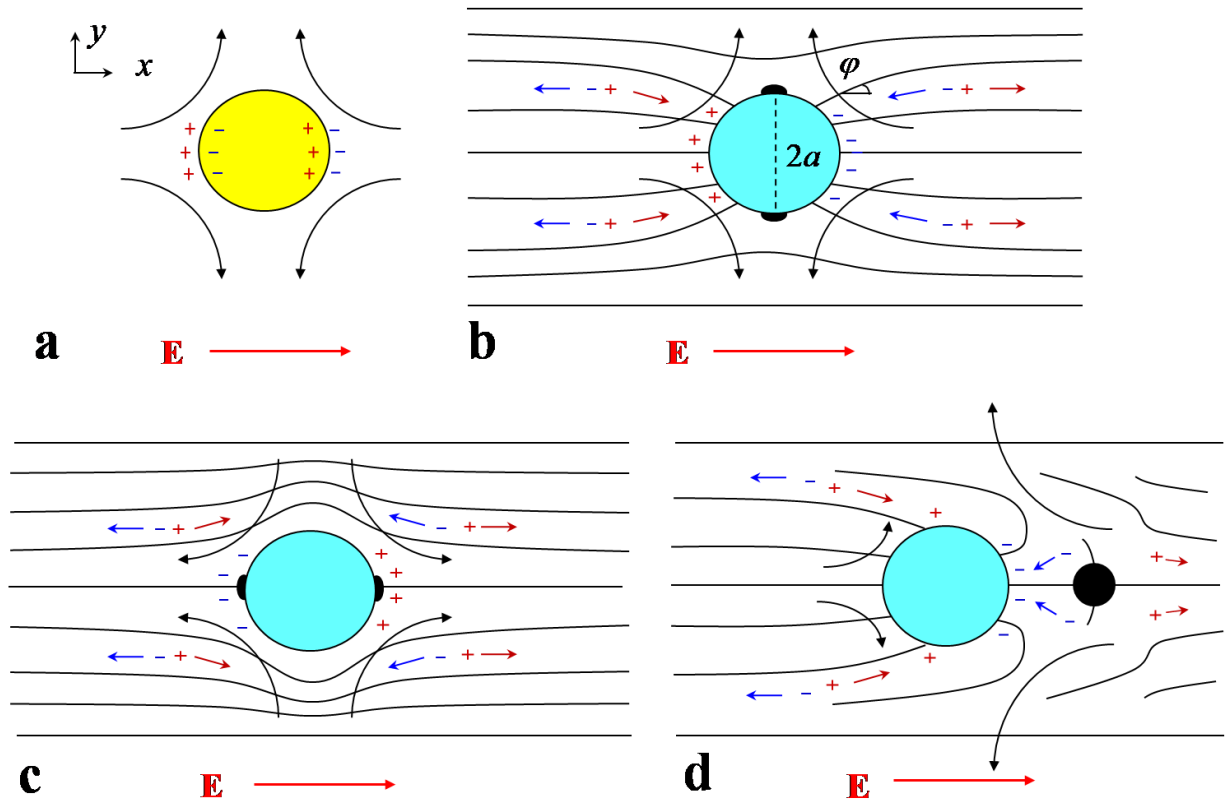

**Figure 1 | Nonlinear electro-osmotic flows in isotropic and LC electrolytes. (a)** ICEO around a conductive particle in water. **(b)** and **(c)** LCEO caused by director distortions around a non-conductive particle with perpendicular **(b)** and tangential **(c)** anchoring. The electric field accumulates ions at the opposite sides of the particle. The separated charges are forced by the electric field into LCEO motion with directions shown by large curved arrows. The LCEO flows are quadrupolar in both **(b)** and **(c)**, but of opposite directionality, of a puller **(b)** and pusher **(c)** type. **(d)** Dipolar symmetry of the director around a perpendicularly anchored particle breaks the fore-aft symmetry of LCEO flows, pumping the LC from right to left.

As easy to deduce from Fig.1b,c, reversing the field polarity reverses the polarity of separated charges, but does not change the polarity of the resulting LCEO flow. The latter implies that the LCEO velocity depends on the square of the field, $u \propto E^2$, and that LCEO can be driven by an AC field, a very useful feature from the point of view of creating steady flows and avoiding chemical reactions at electrodes. One should also expect a linear dependence on the particle's size, $u \propto a$, since the scale of charge separation is determined by the extension of director distortions. In principle, the distortions of the director can be created by means other than the colloidal particles, for example, by surface alignment patterning or by external fields.

To substantiate the qualitative consideration above, let us estimate the density $\rho$ of charges created as a result of conductivity anisotropy in a distorted LC and establish how it depends on the typical scale $a$ of director distortions, $E$, and $\Delta\sigma$. For simplicity, consider two dimensional geometry. We assume that the nematic dielectric anisotropy is small, $\Delta\varepsilon \ll \bar{\varepsilon}$, where $\bar{\varepsilon} = (\varepsilon_\parallel + \varepsilon_\perp)/2$, and that the director distortions are weak, $\hat{\mathbf{n}} = (1, \varphi)$, where $\varphi = \varphi(x, y)$ is a small tilt angle that the director makes with the $x$-axis, Fig.1b. The field, applied along the $x$-axis, creates ionic currents $J_i = \sigma_{ij} E_j$, where $\sigma_{ij} = \sigma_\perp \delta_{ij} + \Delta\sigma n_i n_j$ is the conductivity tensor; $i$ and $j$ stand for $x$ and $y$. For small $\varphi$'s, the current components are $J_x = \sigma_\parallel E_x + \Delta\sigma \varphi E_y$ and $J_y = \sigma_\perp E_y + \Delta\sigma \varphi E_x$; note the field component $E_y$ induced by separation of charges. Using the charge conservation law, $\mathrm{div}\mathbf{J} = 0$ and Poisson's equation $\mathrm{div}\mathbf{D} = \rho$ (where $\mathbf{D}$ is the electric displacement) one obtains the charge density $\rho(x, y)$ caused by conductivity anisotropy $\Delta\sigma$ and director gradients $|\partial\varphi/\partial y| \sim 1/a$:

$$\rho(x, y) = \left( -\frac{\Delta\sigma}{\bar{\sigma}} + \frac{\Delta\varepsilon}{\bar{\varepsilon}} \right) \varepsilon_0 \bar{\varepsilon} E_x \frac{\partial\varphi}{\partial y}; \qquad (1)$$

here $\bar{\sigma} = (\sigma_\parallel + \sigma_\perp)/2$ is the average conductivity. The space charge experiences a bulk force of density $f \propto \rho E_x$ that drives the LCEO flows with a characteristic velocity estimated from the balance of $f$ and viscous resistance $\eta u / a^2$:

$$|u| = \alpha \frac{\varepsilon_0 \bar{\varepsilon}}{\eta} \left| \frac{\Delta \sigma}{\bar{\sigma}} - \frac{\Delta \varepsilon}{\bar{\varepsilon}} \right| a E_x^2. \tag{2}$$

Here the numerical coefficient $\alpha$ on the order of 1 is introduced to account for the approximations such as using $1/a$ as a measure of director gradients and replacing anisotropic viscosity of the LC with its average value $\eta$.

When the quadrupolar symmetry of director distortions is broken, the LCEO flows would acquire dipolar symmetry and thus cause directional pumping, Fig.1d. It is known that spheres with normal boundary conditions in the nematic can produce a dipolar director with a so-called hyperbolic hedgehog, a point defect residing near one of the poles of the sphere, Fig.1d, Ref.[7]. The LCEO flows around such a particle should be fore-aft asymmetric, Fig.1d. Another approach to create a unidirectional pumping would be to create a "Janus anchoring" particle with tangential anchoring on one side and normal anchoring on the other side.

**Liquid crystal materials and visualization of flows.** For experimental demonstration of LCEO, we used dry soda lime glass spheres (purchased from *ThermoScientific*) to create director distortions. The flows were visualized by micro-particle imaging velocimetry ($\mu$PIV)[8,9]. The spheres of a diameter $2a = (50 \pm 2)\,\mu\text{m}$ are placed in a nematic cell of thickness $h = 60\,\mu\text{m}$. To avoid field-induced dielectric torques on the director, we formulated a nematic mixture with zero dielectric anisotropy ($\Delta\varepsilon \leq 10^{-3}$) and average dielectric permittivity $\varepsilon = 6$, by mixing two nematics MLC7026-000 and E7 (both from *EM Industries*) in weight proportion 89.1:10.9. We measured the concentration of ions $n_o \approx 10^{19}\,\text{ions} \cdot \text{m}^{-3}$ and conductivity anisotropy $\Delta\sigma/\bar{\sigma} = 0.3$. The overall director was aligned along the $x$-axis, $\hat{\mathbf{n}}_0 = (1,0,0)$, by a rubbed polymer coating

PI2555 (*Nissan Chemicals*) at top glass plate and rubbed *Norland* film at the bottom glass plate; the latter was used to immobilized the particles.

The LC is doped with a small amount ($\sim 0.01\,\text{wt}\%$) of tracers, representing fluorescent polystyrene spheres (*Bangs Laboratories*) of diameter $2R = 0.19\,\mu\text{m}$. Small size of the tracers as compared to the radius of the spheres allows one to eliminate the potential influence of dielectrophoretic effects [8]. The microscope was focused at the middle plane of the cell where the fluid motion is predominantly horizontal. The tracers cause no visible distortions of the director and are practically non-polarizable. The fluorescent signal of tracers was recorded as a TIFF image with a typical exposure time $\Delta\tau = 325\,\text{ms}$. A superimposition of over 1500 images in one experiment renders a single composite picture showing the streamlines of the flows around the spheres. The experimental flow velocity vector fields were obtained using μPIV software [9] which correlates the position of tracers in consecutive images.

Untreated spheres yield tangential anchoring of $\hat{\mathbf{n}}$, with the director configuration as in Fig.1c, 2c. The spheres treated with octadecyl-dimethyl(3-trimethoxysilylpropyl) ammonium chloride (DMOAP) produce perpendicular alignment and two types of distortions, stable in cells of thickness close to $2a$, Ref. [10]: (A) quadrupolar "Saturn ring" configuration with an equatorial disclination ring, Fig.1b, 2a, Ref. [11], and (B) dipolar configuration with a hyperbolic hedgehog on one side of the sphere [7], Fig.1d, 2e. LCEO is driven by an in-plane AC electric field of frequency 5 Hz applied along the $x$-axis using two aluminum electrodes separated by a gap $L =$ 10 mm. All experiments were performed at room temperature 22$^\text{o}$C.

**Liquid crystal enabled electro-osmosis and pumping.** Experimental demonstration of LCEO, visualized by μPIV, is presented in Fig.2. The quadrupolar director patterns, Fig.2a,c, produce quadrupolar LCEO flows. As expected from Fig.1b,c, different surface anchoring of the spheres (with opposite signs of director gradients in Eq.(1)) leads to opposite polarities of the flows, of the "puller" type for the normal anchoring, Fig.2b, and the "pusher" type for the tangential anchoring, Fig. 2d. The dipolar director pattern, Fig.2e, produces an LCEO with broken fore-aft symmetry, Fig.2f. The LC is pumped from right to left in Fig.2f. Pumping is powered by two strong vortices near the point defect-hedgehog, Fig. 2e,f.

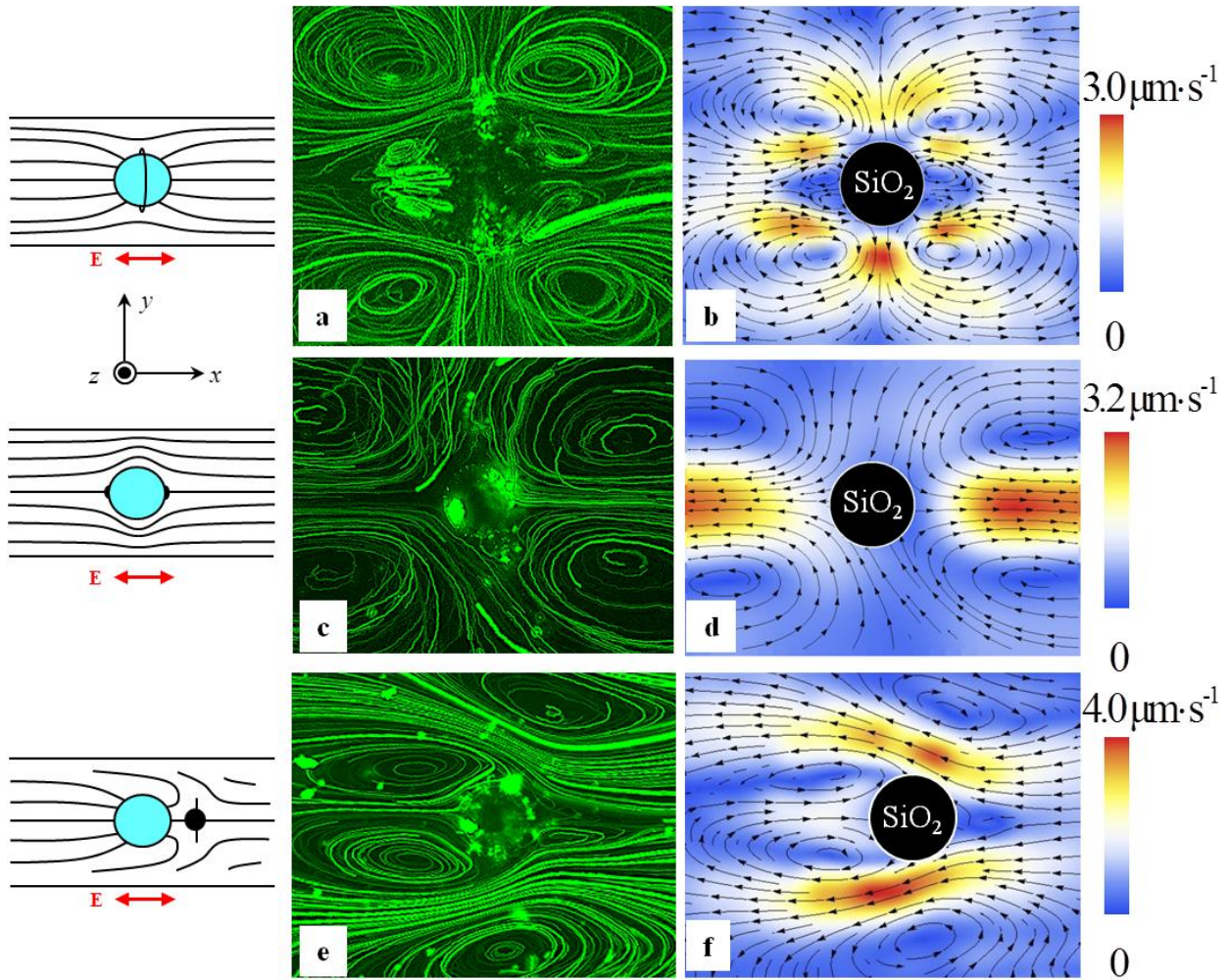

**Figure 2 | Streamlines and velocity fields of LCEO around glass spheres. (a),(b)** perpendicular anchoring with quadrupolar director patterns; **(c),(d)** bipolar sphere with tangential anchoring. Velocities in **(b)** and **(d)** are of opposite polarity, creating patterns of a puller **(b)** and pusher **(d)** type. **(e),(f)** sphere with perpendicular anchoring and dipolar director pattern; note

pumping from right to left in (**f**). Particle diameter $2a = 50\,\mu m$, AC electric field $E = 26\,mV \cdot \mu m^{-1}$.

The amplitude $|u^{max}|$ of LCEO velocities is linearly proportional to $a$, Fig.3a, and to $E^2$, Fig.3b, as expected from Eq.(2). These amplitudes $|u^{max}| \approx (3-4)\,\mu m \cdot s^{-1}$ achieved in the nematic LC at $E = 26\,mV \cdot \mu m^{-1}$ are much higher than the ICEO velocities around dielectric spheres in an isotropic electrolyte. To make a quantitative comparison, we explored ICEO flows around the same (untreated) glass spheres but placed in aqueous electrolyte containing 1 mM of KCl. In this case $u_{diel}$ does not exceed $0.1\,\mu m \cdot s^{-1}$ even at a high field $E = 70\,mV \cdot \mu m^{-1}$. For a comparable electric field driving, the ratio of LCEO and ICEO velocities is on the order of $\frac{|u^{max}|}{|u_{diel}|} \approx \frac{\varepsilon_{LC}\eta_{water}a}{\varepsilon_{water}\eta_{LC}\lambda_D}$, where the subscripts refer to the medium. Despite the fact that $\varepsilon_{LC}/\varepsilon_{water} \sim 0.1$ and $\eta_{water}/\eta_{LC} \sim 0.1$, it is the large ratio $a/\lambda_D \sim 10^4$ that makes the LCEO velocities about two order of magnitude higher as compared to ICEO velocities, thanks to the separation of charges over the scales $\sim a$. The theoretical estimate, Eq.(2), with $\alpha = 1$, $\varepsilon_{LC} = 6$, $\eta_{LC} \approx 0.07\,kg \cdot m^{-1} \cdot s^{-1}$, $a \approx 25\,\mu m$, $E = 26\,mV \cdot \mu m^{-1}$, predicts a characteristic velocity $|u^{max}| \approx 4.5\,\mu m \cdot s^{-1}$, close to the experimental $|u^{max}|$, Fig.3a,b. It is hard to expect a better agreement since the proper theoretical analysis should include factors such as anisotropic viscoelasticity of the LC, presence of bounding substrates, etc. Note that the actual director gradients might be weaker than $\sim 1/a$ because of finite strength of surface anchoring.

Pumping around spheres with the dipolar director is illustrated by the behavior of volumetric flow $Q_x(x) = \frac{2}{3} h \int_{-y_0}^{y_0} u_x(x,y) dy$ passing through the $xz$ cross-sections of the cell as a function of $x$-coordinate in the range $|x| \leq x_0$, Fig.3c. Here $x_0 = y_0 = 150\,\mu m$; the functions $u_x(x,y)$ and $u_y(x,y)$ are known from the experiment, Fig.2f. The fore-aft asymmetry in Fig. 2e,f results in pumping of the LC from right to left, as $Q_x(x)$ is everywhere negative. In the orthogonal $y$-direction, there is only mixing, but no pumping, as the corresponding flow function $Q_y(y) = \frac{2}{3} h \int_{-x_0}^{x_0} u_y(x,y) dx$ is antisymmetric.

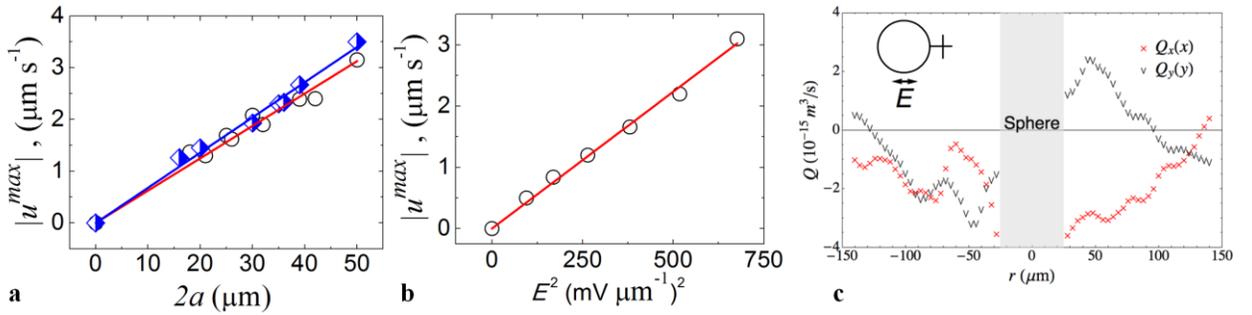

**Figure 3 | Quantitative parameters of LCEO around glass spheres**. **(a)** Maximum LCEO velocity around a tangentially anchored (circles) and normally anchored (diamonds) spheres increases linearly with the diameter $2a$; $E = 26\,mV\cdot\mu m^{-1}$; in both cases the director pattern is quadrupolar. **(b)** Maximum LCEO velocity around a tangentially anchored sphere of diameter $2a = 50\,\mu m$ grows as $E^2$. **(c)** Volume of LC flow around a sphere with a hedgehog ($2a = 50\,\mu m$), Fig.2f, passing along the $x$ ($Q_x(x)$) and $y$ ($Q_y(y)$) axes. The LC is pumped from right to left.

**Spatial dependencies of electro-osmotic velocities.** The LCEO flows follow certain angular and radial dependencies, as established by Fourier analysis of the velocity fields around normally anchored spheres, Fig.4. The radial $u_r(r,\theta)$ and azimuthal $u_\theta(r,\theta)$ components are represented as series of sines and cosines with the Fourier coefficients that depend on the radial distance $r$ from the center of the sphere:

$$u_i(r,\theta) = u_i^{(0)} + \sum_{n=1}^{\infty}\left[u_{ic}^{(n)}(r)\cos n\theta + u_{is}^{(n)}(r)\sin n\theta\right], \tag{3}$$

where $i$ stands for $r$ or $\theta$, denoting the radial and azimuthal components of the velocity, respectively; $\theta$ is the polar angle measured with respect to the $x$-axis.

For spheres with the quadrupolar director, the flows are symmetric with respect to the planes $xz$ and $yz$ that cross the sphere's center, Fig.2a-d. The vortices are elongated along the overall director $\hat{\mathbf{n}}_0 = (1,0,0)$, which can be attributed to anisotropy of conductivity $\Delta\sigma/\bar{\sigma} = 0.3$ and viscosity $\eta_\parallel/\eta_\perp = (0.5 - 0.7)$. To account for these anisotropies, Fourier description of Fig.2b requires three even harmonics ($n = 2, 4, 6$), Fig.4a,b; these harmonics produce a very close match of the experiment, compare Fig.4c to Fig.2b. Far from the sphere, the coefficients $u_{rc}^{(2)}(r) = u_{\theta s}^{(2)}(r)$, $u_{rc}^{(4)}(r) = 2u_{\theta s}^{(4)}(r)$ and $u_{rc}^{(6)}(r) = 3u_{\theta s}^{(6)}(r)$ show dependencies $\propto 1/r^3$ and satisfy the incompressibility condition.

For the dipolar configuration, Fig. 2e,f, the LCEO pattern contains odd harmonics that are needed in order to describe pumping of the LC from the hedgehog towards the sphere, Fig. 4f, Fig.3c. The coefficients of the first harmonic $n = 1$ decay with the distance as $u_{rc}^{(1)}(r) = u_{\theta s}^{(1)}(r) \propto 1/r^2$, while the coefficients with $n = 2$ to 6 decay faster, $\propto 1/r^3$, satisfying the incompressibility condition. Note that the pumping effect around immobilized spheres with a satellite hedgehog defect is a counterpart of the nonlinear electrophoresis of dielectric spheres freely suspended in a nematic and driven by an AC field [12, 13]. Both effects are caused by the conductivity anisotropy of the LC and broken fore-aft symmetry of director distortions.

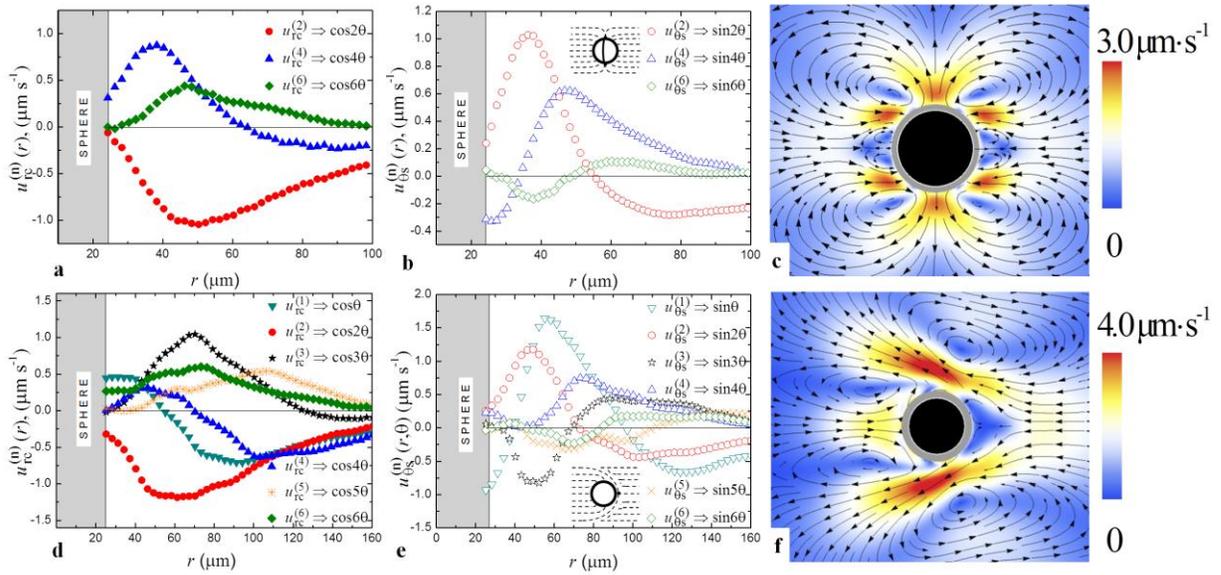

**Figure 4 | Fourier analysis of velocity for quadrupolar (top) and dipolar (bottom) distortions. (a)** Radial $u_{rc}^{(n)}(r)$ and **(b)** azimuthal $u_{\theta s}^{(n)}(r)$ coefficients for LCEO flows triggered by a quadrupolar director pattern, Fig.2b; **(c)** velocity field reconstructed with the three harmonics that show a close similarity to the experimental data in Fig.2b; **(d),(e),(f),** the same, but for a sphere with a dipolar director and LCEO pumping, Fig.2f; compare to Fig.2f. Particle diameter $2a = 50\,\mu m$.

## Discussion

To summarize, the intrinsic conductivity anisotropy of LCs leads to a rich variety of electrokinetic effects, as the electric field applied to a distorted LC creates space charges that are driven by this field, which results in LCEO flows with velocities that are quadratic in the field. The latter allows one to use an AC field driving. The LCEO flows around poorly polarizable particles, such as glass spheres, are of significant amplitude, not reachable in the case of isotropic electrolytes. The reason is that the charge separation occurs on the length scales of director distortions. By controlling the director patterns, one can dramatically change the nature of LCEO flows, for example, trigger a pumping effect by changing the pattern from quadrupolar to dipolar. In the demonstrated mechanism of LCEO, the particles are required only to create the director distortions. The mechanism is applicable to particle-free samples, provided the director distortions are created by some other means, such as patterned surface alignment or external fields. The LCEO thus opens a broad spectrum of potential lab-on-a-chip, optofluidic and microfluidic applications, as the director distortions in LCs are known to be easily triggered and modified statically and dynamically, by a variety of factors, from temperature and electromagnetic fields, to light-induced reorientation in the bulk and photo-alignment at the surfaces. Recently, a rich potential of the LCs as microfluidic media has been demonstrated in pressure-driven effects [14, 15]. The described effect of LCEO adds a new non-mechanical LC-based approach to the design and engineering of lab-on-a-chip systems.

## Methods

**Ionic mobilities.** We obtained the ionic mobility parallel and perpendicular to the director $\hat{\mathbf{n}}$, by measuring the transient current peak when a squared wave voltage of 1Hz was applied through the liquid crystal cell [16]. The liquid crystal mixture was injected into commercially available E.H.C. Co glass cells that consist of two pieces of glass with transparent indium tin oxide (ITO) patterned on a square area 1x1cm$^2$. We used two types of cells, one that produces a planar alignment of the director and another one that produces homeotropic alignment. The thickness of the cells was $h = 15 \mu m$. The ionic mobility is determined from the time $\tau_p$ required by the ions to migrate through the cell. The quantity $1/U\tau_p$ was measured to be constant over the range of applied voltages, $U = (4\text{-}7)\text{V}$, with values $1/U\tau_p = 5.5\,\text{V}^{-1}\text{s}^{-1}$ for the planar cell and $1/U\tau_p = 7.75\,\text{V}^{-1}\text{s}^{-1}$ for the homeotropic one. The anisotropic mobility ratio is $\mu_\parallel/\mu_\perp = 1.4 \pm 0.1$; the corresponding ratio characterizing anisotropy of conductivity is $\Delta\sigma/\bar{\sigma} = 2(\sigma_\parallel - \sigma_\perp)/(\sigma_\parallel + \sigma_\perp) \approx 0.3$.

**Anisotropy of viscosity.** To determine the viscosity anisotropy of the LC, we explored Brownian diffusion of $2a = 5.0\,\mu\text{m}$ glass spheres dispersed in the LC bulk. Untreated spheres produced tangential anchoring; spheres treated with DMOAP produced normal anchoring. In cells with planar alignment, the spheres display anisotropic Brownian motion with two self-diffusion coefficients [17]: $D_{\parallel/\perp} = k_B T / 6\pi\eta_{\parallel/\perp} a$; here $k_B$ is the Boltzmann constant, $T$ is the temperature; $\eta_\parallel$ and $\eta_\perp$ are the viscosity coefficients corresponding to the motion parallel and perpendicular to $\hat{\mathbf{n}}$, respectively. The two independent viscosity coefficients $\eta_\parallel$ and $\eta_\perp$ are determined by tracking the position of the particles at a given time and measuring the mean square displacement (MSD) vs time, $\langle\Delta x^2\rangle = \langle(x(t+\tau) - x(t))^2\rangle = 2D_\parallel \tau$, $\langle\Delta y^2\rangle = \langle(y(t+\tau) - y(t))^2\rangle = 2D_\perp \tau$ along and perpendicular to the director, respectively. The time lag between images was $\tau = 100\,\text{ms}$, long enough to avoid anomalous diffusion in liquid crystals [18]. For spheres with tangential anchoring the diffusion anisotropy is $D_\parallel/D_\perp = 2.07 \pm 0.05$ and viscosity coefficients $\eta_\parallel = 40\,\text{mPa}\cdot\text{s}$ .and $\eta_\perp = 83\,\text{mPa}\cdot\text{s}$, while for normal anchoring $D_\parallel/D_\perp = 1.44 \pm 0.05$ and viscosity coefficients $\eta_\parallel = 54\,\text{mPa}\cdot\text{s}$ .and $\eta_\perp = 78\,\text{mPa}\cdot\text{s}$.

**Nonuniformity of the electric field.** Since the electric field is applied in the plane of the cell by two aluminum foils, its distribution is generally nonuniform. To calculate the field configuration, a commercial finite-element modelling simulator, COMSOL Multiphysics, was used. The experimentally known parameters are the distance between the electrodes $L = 10$ mm, cell thickness $h = 60\,\mu m$, dielectric permittivity of glass 3.9 and of liquid crystal 6. As shown in Fig.5, in the central part of the cell, the field is uniform and parallel to the glass plates. The experiments were performed for this central part where the field can be assumed uniform. The field in the center of cell is calculated from the known applied voltage $U$ and $L$ as $E = \beta U / L$, where $\beta$ is a correction factor obtained by numerical simulations to be 0.65 for the experimental cells used.

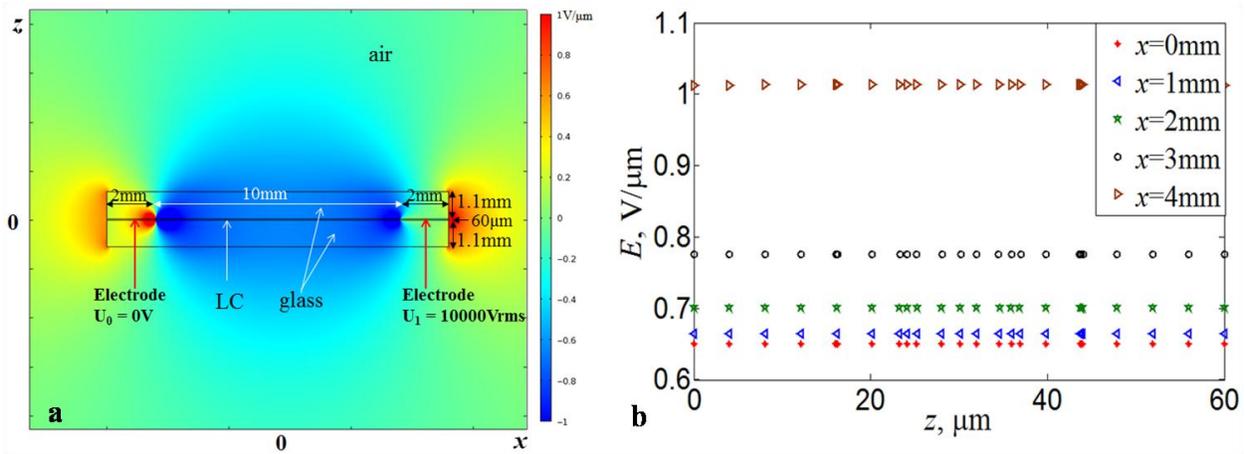

**Figure 5. | Electric field in the nematic cell with two aluminum foil electrodes.** (**a**) Electric field pattern; (**b**) Dependencies of the local field on the location within the nematic slab. Cell thickness $h = 60\,\mu m$. Thickness of each glass plates 1.1 mm.

**References**

1. Ramos A. Electrohydrodynamic Pumping in Microsystems. In: *Electrokinetics and Electrohydrodynamics in Microsystems* (eds Ramos A). Springer Vienna (2011).

**Acknowledgements**

This work was supported by NSF grant DMR 1104850. The authors declare no competing financial interests.

**Authors contributions**

I.L. performed μPIV studies of electro-osmotic flows around spheres with normal anchoring, C.P. explored electro-osmotic flows around spheres with tangential anchoring, J.X. established electric field strength in the nematic cell through numerical simulations, I.L., C.P., J.X. S.V.S. and O.D.L. analyzed the data, O.D.L. directed the research and wrote the text with an input from all co-authors.